\documentstyle[12pt]{article}

\large
\newcommand{\be}{\begin{eqnarray}}
\newcommand{\ee}{\end{eqnarray}}
\newcommand\del{\partial}

\newcommand\half{\frac 12}

\begin{document}
\setlength{\baselineskip}{21pt}
\pagestyle{empty}
\vfill
\eject
\begin{flushright}
SUNY-NTG-94/2
\end{flushright}

\vskip 2.0cm
\centerline{\bf The spectrum of the Dirac operator near zero virtuality}
\vskip 0.2 cm
\centerline{\bf for $N_c = 2$}
\vskip 2.0 cm
\centerline{Jacobus Verbaarschot}
\vskip .2cm
\centerline{Department of Physics}
\centerline{SUNY, Stony Brook, New York 11794}
\vskip 2cm

\centerline{\bf Abstract}
We study the spectrum of the QCD Dirac operator near zero virtuality
for $N_c =2$. According to a universality argument, it can be described
by a random matrix theory with the chiral structure of QCD, but with $real$
matrix elements.

Using results derived by Mehta and Mahoux and Nagao and Wadati, we are
able to obtain an analytical result for the microscopic spectral density
that in turn is the generating function for Leutwyler-Smilga type
spectral sum rules.

\vfill
\noindent
\begin{flushleft}
SUNY-NTG-94/2\\
January 1994
\end{flushleft}
\eject
\pagestyle{plain}

\vskip 1.5cm
\renewcommand{\theequation}{1.\arabic{equation}}
\setcounter{equation}{0}
\noindent
\centerline{\bf 1. Introduction}
\vskip 0.5cm
Quantum Chromodynamics (QCD) for two colors ($N_c=2$)
shares many common features
with QCD for $N_c = 3$, generally believed to be the
correct theory of strong interactions. However,
there are important differences. In particular, the low-energy excitations
do not only involve mesons but also baryons,
which consist out of two quarks and are bosons.
The corresponding effective theory \cite{DIAKONOV-PETROV-1993}
has a much richer
structure with many Goldstone particles than for $N_c = 3$.
For example, for $N_f= 2$ we have nine Goldstone bosons
\cite{DIAKONOV-PETROV-1993} for the
$SU(4) \rightarrow SO(4)$ symmetry breaking scheme, instead of the usual 3 (see
\cite{VYSOTSKII-ETAL-1985,DIMOPOULOS-1980,PESKIN-1980}
for a detailed discussion of the different
chiral symmetry breaking schemes for $N_c = 2$).

As we have learnt from the work of Leutwyler and Smilga
\cite{LEUTWYLER-SMILGA-1993}, the existence
of a low-energy effective theory imposes severe restrictions on the spectrum of
the Dirac operator in the form of sum rules for the inverse powers of its
eigenvalues.
Recently, we have shown
\cite{SHURYAK-VERBAARSCHOT-1993,VERBAARSCHOT-ZAHED-1993}
that the spectral sum rules can also be obtained
from a random matrix theory with the chiral symmetry of the
Dirac operator. This theory enabled us to derive the joint spectral
density that generates $all$ sum rules obtained by Leutwyler and Smilga.
In fact, arguments from random matrix theory imply that
the spectral correlations {\it near zero virtuality} are uniquely determined
by the symmetries of the system
\cite{SHURYAK-VERBAARSCHOT-1993,VERBAARSCHOT-ZAHED-1993}. In other words
they are universal (see \cite{BREZIN-ZEE-1993} for a systematic study of this
issue). This point has been known for quite some time in the study of spectra
of classically chaotic systems
\cite{BOHIGAS-GIANNONI-1984,SELIGMAN-VERBAARSCHOT-ZIRNBAUER-1984}
\cite{SIMONS-SZAFER-ALTSCHULER-1993} and
the theory of $S$-matrix fluctuations
\cite{VERBAARSCHOT-WEIDENMUELLER-ZIRNBAUER-1985}.
For example, both Ericson fluctuations \cite{ERICSON-1960}
and universal conductance fluctuations
\cite{LEE-STONE-FUKUYAMA-1987,SLEVIN-NAGAO-1993} can be unified
within the latter context (see \cite{WEIDENMUELLER-1991}
for a discussion of this similarity).
In particular, this means that the microscopic spectral density,
defined as the
$V_4\rightarrow\infty$ limit of the spectral density while the
eigenvalues are rescaled $\sim V_4$, is a universal function.

The random matrix theory that corresponds to the standard scheme of chiral
symmetry breaking, has $complex$ matrix elements. For this reason
we have called it the chiral unitary ensemble (chGUE). In the framework
of this model, new  sum rules \cite{VERBAARSCHOT-1994S}
can be derived with the help of the Selberg \cite{SELBERG-1944,MEHTA-1991}
integral formula.

Because the effective theory
for $N_c = 2$ is different from the generic case, we
expect a different microscopic spectral density and different sum rules.
This raises the question what is the correct random matrix theory for
$N_c = 2$. The answer becomes clear if one considers the matrix elements
of the Dirac operator. In this case it is possible to choose a basis in which
they are real, as opposed to three and more colors where they are complex.
This reminds us of the three universality classes in random matrix
theory \cite{DYSON-1962}, the Gaussian Orthogonal Ensemble (GOE),
the Gaussian Unitary Ensemble (GUE) and the Gaussian
Symplectic Ensemble (GSE). The first correspond the real symmetric matrices,
the
second to Hermitean complex matrices and the third to quaternion matrices.
It is clear that the correct random matrix ensemble should not only embody
the chiral symmetry of the Dirac operator but also satisfy the additional
constraint that the matrix elements are real.
{}From now on we will call this matrix ensemble the chiral orthogonal ensemble
(chGOE).
The third possibility is realized for fermions in the adjoint
representation \cite{VERBAARSCHOT-1994A}. In this case it is possible to
regroup
the matrix elements of the Dirac operator in terms of quaternions. The
corresponding ensemble will called the chiral symplectic ensemble
(chGSE).

In a separate publication \cite{VERBAARSCHOT-1994I}
we will show that the chGOE describes the
spectrum of the Dirac operator in a liquid of instantons
\cite{SHURYAK-VERBAARSCHOT-1990}, whereas it
is not given by the chGUE. In this model the Dirac operator is diagonalized
in the space of fermionic zero modes. Indeed, the matrix elements of the Dirac
operator are real. Also the corresponding sum rules will be given
elsewhere \cite{VERBAARSCHOT-1994S}.

In this work we report on the calculation of the microscopic spectral density
(defined in section 2) of the Dirac operator for $N_c = 2$.
The argument that the matrix elements of the Dirac operator are real
is presented in section 3. The corresponding random matrix theory
and the joint eigenvalue distribution is derived in section 4. The exact
spectral
density for any finite size random matrix is calculated in section 5.
The microscopic spectral density is obtained in section 6 and concluding
remarks are made in section 7. Some technical details are worked out in three
appendices.

\vskip 1.5cm
\renewcommand{\theequation}{2.\arabic{equation}}
\setcounter{equation}{0}
\centerline{\bf 2. Formulation of the problem}
\vskip 0.5 cm

The distribution of the eigenvalues of the Dirac operator are determined
by the fluctuations of the gauge field which are subject to the Euclidean
QCD partition function
\be
Z = \sum_\nu e^{i\nu\theta} \langle \prod^{N_f}_{f=1}\prod_{\lambda_n>0}
(\lambda_n^2 + m_f^2) m_f^{\nu}\rangle_{S_\nu(A)},
\ee
where the average $\langle\cdots\rangle_{S_\nu(A)}$
is over gauge field configurations
with topological quantum number $\nu$ weighted by the gauge field action
$S_\nu(A)$. The product is over all eigenvalues of the
Dirac operator, and relevant observables are obtained by differentiation
with respect to the masses $m_f$. The number of flavors is denoted by $N_f$.
The  factor $\exp i\theta\nu$ represents the
topological term in the action.

The condensate can be expressed as a derivative of the partition function
\be
\langle \bar q q \rangle = \lim_{m_f\rightarrow 0} \lim_{V\rightarrow \infty}
\frac i{V_4}
\frac d{dm_f} \log Z(m_f).
\ee
According to the Banks-Casher \cite{BANKS-CASHER-1980} formula we have
\be
\langle \bar q q \rangle = i \pi \frac{\langle\rho(0)\rangle}
{V_4},
\ee
where the spectral density $\rho(\lambda)$ is defined as
\be
\rho(\lambda) = \sum_{\lambda_n} \delta(\lambda -\lambda_n).
\ee
It is now clear that, in order to obtain a nonzero value of
$\langle\bar q q \rangle$, we should have
\be
\langle\rho(0)\rangle \sim V_4,
\ee
or, put differently, the spacing between the eigenvalues near zero virtuality
is $\sim 1/V_4$ (Note that for a non-interacting system the spacing between
the eigenvalues is $\sim 1/V^{1/4}$).

As was observed by Leutwyler and Smilga \cite{LEUTWYLER-SMILGA-1993},
(2.5) implies the existence
of a family of new sum rules. The simplest one involves the sum
\be
\frac 1{V^2_4} \sum_{\lambda_n > 0} \left <\frac 1{\lambda^2_n}\right >_\nu
\ee
which should converge to a finite limit for $V_4 \rightarrow \infty$.

The above mentioned sum-rules can be expressed in the microscopic spectral
density defined by
\be
\rho_S(x) = \lim_{ V_4\rightarrow \infty} \frac 1V_4 \left
<\rho(\frac xV_4)\right >_\nu
\ee
in the sector of topological charge $\nu$.
For the  sum (2.6) we find
\be
\int dx \frac{ \rho_S(x)}{x^2}.
\ee
In this paper we will obtain
an analytical expression for
$\rho_S(x)$ for $N_f$ flavors and arbitrary topological charge $\nu$.

\vskip 1.5cm
\renewcommand{\theequation}{3.\arabic{equation}}
\setcounter{equation}{0}
\centerline{\bf 3. Symmetries of the Dirac operator for $N_c =2$}
\vskip 0.5 cm

In this paper we study the Euclidean Dirac operator
\be
D \equiv i\gamma \del + \gamma A,
\ee
where $A$ is an $SU(2)$ valued gauge field. The spectrum of $D$ is defined
by the eigenvalue equation
\be
D \phi_\lambda = \lambda \phi_\lambda.
\ee
The Dirac operator for $SU(2)$ has two symmetries. First, the chiral symmetry,
which is present for any $SU(N_c)$-valued gauge group,
\be
\{\gamma_5, D\} = 0,
\ee
and second, a symmetry specific to $SU(2)$:
\be
[C^{-1} \tau_2 K, D] =0,
\ee
where $C$ is the charge conjugation operator ($\gamma_{\mu}^* = - C\gamma_mu
C^{-1}$ and $C C^* = -1$), and $K$ denotes the
complex conjugation operator. Because of the
first symmetry, the eigenvalues occur in pairs $\pm \lambda$.
The second
symmetry operator has the property that
\be
(C^{-1} \tau_2 K)^2 = 1.
\ee
As is well known from the analysis of the of the time-reversal operator
in random matrix theory (see for example refs. \cite{PORTER-1965,MEHTA-1991}),
this property allows us to choose a basis in which
the Dirac operator is real. In a chiral basis the Dirac operator
therefore has the following general structure
\be
\left ( \begin{array}{cc}
             0 & T \\ \tilde T & 0
\end{array} \right ),
\ee
where $T$ is a real matrix. This is the only information of the Dirac operator
that will be injected into the random matrix model to be defined in section 4.

\vskip 1.5cm
\renewcommand{\theequation}{4.\arabic{equation}}
\setcounter{equation}{0}
\centerline{\bf 4. The chiral random matrix model}
\vskip 0.5 cm

As has been shown in previous section, the matrix elements of the Dirac
operator can be chosen real. The corresponding random matrix theory
with the chiral structure of QCD is defined by the partition function
\be
Z_\nu = \int {\cal D}T P(T)\prod_f^{N_f}\det \left (
\begin{array}{cc} m_f & iT\\
                 i\tilde T & m_f
\end{array} \right ).
\ee
This model is defined for $N_f$ flavors with masses $m_f$ in the chiral limit
($m_f\rightarrow 0$) and in the sector with topological charge $\nu$.
The latter property is implemented by choosing $T$ an
$m\times n$ matrix (for definiteness, $m\ge n$) with $|m-n| =\nu$. With
this choice the matrix of which the determinant is calculated has exactly
$\nu$ zero eigenvalues for $m_f = 0$.
The integral is over all matrix elements of $T$,
$i.e.$, ${\cal D}T$ is the Haar measure. As follows form the
the maximum entropy principle \cite{BALIAN-1968}
the distribution function of the overlap matrix
elements $P(T)$ is chosen Gaussian
\be
P(T) = \exp(-\frac{n\beta}{2\sigma^2} {\rm Tr }T \tilde T).
\ee
In the present case ($T$ real) the value of $\beta =1$, whereas
for complex matrix elements $\beta=2$. With this convention,
a mean-field argument shows that the average level
density for the chGOE is the same as for the chGUE (see appendix A).
One can easily show that the eigenvalues of a matrix with this block
structure occur in pairs $\pm \lambda$, a property that is well-known
for the nonzero eigenvalues of the massless Dirac operator.
The density of modes $N/V_4$ ($N=m+n$) is taken equal to one,
which allows us
to identify $N$ with the volume of space time $V_4$, which we will do from now
on. In agreement with general QCD relations we will always assume
that $\nu \ll N$.

The matrix $T$ can be diagonalized by an $n\times n$ orthogonal matrix
${\cal O}_1$ and an $m\times m$ orthogonal matrix ${\cal O}_2$:
\be
T = {\cal O}_1 \Lambda {\cal O}_2.
\ee
Here, $\Lambda$ is an $n\times m$ diagonal matrix with diagonal
matrix elements $\lambda_k$.
The joint eigenvalue distribution is obtained
immediately by using $\Lambda$,  ${\cal O}_1$
and ${\cal O}_2$ as new integration variables.
Because the integrand only depends on $\Lambda$, the integration over the
orthogonal matrices can be performed trivially. The Jacobian corresponding
to the transformation (4.3) is given by (see appendix B)
\be
J(\Lambda) = \prod_{k<l}^n |\lambda_k^2 -\lambda_l^2|\prod_{k=1}^n
\lambda_{k}^\nu
\ee
The joint probability density of the nonzero eigenvalues is therefore given by

\be
\rho(\lambda_1, \cdots, \lambda_n) = \prod_{k<l}^n| \lambda_k^2 -\lambda_l^2|
                                     \prod_k ^n\lambda_k^{2N_f+\nu} \exp(
{-\frac {n\beta}{2\sigma^2} \sum \lambda_k^2}).
\ee
Since this distribution is symmetric in all eigenvalues, the spectral density
is simply obtained by integrating it over all eigenvalues except for one:
\be
\rho(\lambda_1) = \frac{\int d\lambda_2 \cdots d\lambda_n
\rho(\lambda_1,\cdots, \lambda_n)}{\int d\lambda_1 \cdots d\lambda_n
\rho(\lambda_1,\cdots, \lambda_n)}.
\ee
The normalization integral in the denominator will be denoted by $Z(n)$.

\vskip 1.5cm
\renewcommand{\theequation}{5.\arabic{equation}}
\setcounter{equation}{0}
\centerline{\bf 5. Calculation of the spectral density}
\vskip 0.5 cm
To make contact with  what is also
called the  orthogonal generalized Laguerre ensemble, we introduce new
integration variables by
\be
x_k = \frac {n\lambda_k^2}{2\sigma^2}.
\ee
If we absorb the constants in a redefinition of the normalization constant $Z$,
the spectral density is given by
\be
\rho(\lambda_1) d\lambda_1 =\rho_L(x_1)dx_1
= \frac {dx_1}{Z} \int_0^\infty dx_2 \cdots dx_n
\prod_{k<l}^n |x_k - x_l| \prod_{k=1}^n
x_k^{a}\, \exp({-\sum_{k=1}^n x_k}).
\ee
For convenience we have introduce the notation
\be
a = N_f -\frac 12 +\frac {\nu}2.
\ee

General expressions for spectral densities given by
these types of integrals have been derived by
Mahoux and Mehta \cite{MAHOUX-MEHTA-1991}. The result
is expressed in skew-orthogonal polynomials $R_k$
defined by
\be
\langle R_{2k}, R_{2l+1} \rangle &=& r_k \delta_{kl},\nonumber\\
\langle R_{2k}, R_{2l} \rangle &=&
 \langle R_{2k+1}, R_{2l+1} \rangle =0,
\ee
with the scalar product given by
\be
\langle f, g \rangle = \int_0^\infty dx x^a e^{-x}
\int_0^\infty dy y^a e^{-y} \epsilon(x-y) f(x) g(y),
\ee
where  $\epsilon(x-y) = \frac 12 {\rm sign}(x-y)$.
For this weight function the normalization constants $r_k$ are known
explicitly
\be
r_k = h_{2m}^{2a+1} = \frac{\Gamma(k+1)\Gamma(2a+k+2)}{2^{2a+2k+2}},
\ee
where $h_{2m}^{2a+1}$ are the normalization constants of the monic
generalized Laguerre polynomials with index $2a+1$.
They are fixed by the normalization integrals $Z(n)$ (see below (4.6)).

The result for the spectral density is given by
\be
\rho_L(x) = \sum _{m=0}^{(n/2)-1} \frac 1{ r_m} (\phi_{2m}(x)\phi'_{2m+1}(x)-
\phi_{2m+1}(x)\phi'_{2m}(x)),
\ee
where
\be
\phi_{k}(x) = \int_0^\infty dy y^a e^{-y} \epsilon(x-y) R_k(y).
\ee
{}From the definition of the skew-symmetric scalar product one finds for
the normalization $\int_0^\infty \rho_L(x) dx = n$.

The polynomials $R_k$ can be obtained most conveniently by expanding them
in the monic generalized Laguerre polynomials
$C_n^{2a}(x) \equiv  n! L^{2a}_n(2x)/(-2)^n$:
\be
R_{2m}(x) &=& \sum_{k=0}^{2m} a_{m\, k} C_{2m-k}^{2a}(x),\nonumber\\
R_{2m+1}(x) &=& \sum_{k=0}^{2m+1} b_{m\, k} C_{2m+1-k}^{2a}(x).
\ee
Starting from the identity
\be
\langle (\frac ax -1) f, g\rangle - \langle f', g\rangle =
- \int_0^\infty x^{2a} e^{-2x} f(x) g(x) dx
\ee
Nagao and Wadati \cite{NAGAO-WADATI-1991}
where able to derive recursion relations for the
expansion coefficients
\be
a_{m\, n+1} &=& -\half(2m-n) a_{m\, n}, \qquad n \ge 0,\nonumber\\
b_{m\, n+1} &=& -\half(2m-n+1) b_{m\, n}, \qquad n \ge 2,\nonumber\\
b_{m\, 2} &=& -\frac m2(2b_{m\,1}+2a+2m+1),
\ee
where $b_{m\,1}$ is not determined by the recursion relation, and in fact,
does not contribution the spectral correlation functions. The intial
conditions are fixed by
\be
a_{m\,0}=b_{m\,0}= 1.
\ee
The solution of these recursion relations is straightforward:
\be
a_{m\, n} &=& \frac 1{(-2)^n} \left (\begin{array}{c}  2m\\ n
\end{array} \right) n!,\\
b_{m\, n} &=& \frac {2a+2m+1}{(-2)^{n-1}} \left (\begin{array}{c}  2m\\ n -1
\end{array}\right)  (n-1)!,\qquad n \ge 2
\ee
where we made the choice (does $not$ satisfy (5.14) for $n=1$)
\be
b_{m\, 1} = \half(2a +2m + 1).
\ee

Remarkably, using the identity $\sum_{k=0}^m L_{m-k}^\alpha = L^{\alpha+1}_m$
(see \cite{GRADSHTEYN-RYZHIK-1980})
the sums in eq. (5.9) can be performed exactly. The result is
\be
R_{2m}(x) &=& C_{2m}^{2a+1}(x),\nonumber\\
R_{2m+1}(x) &=& C^{2a}_{2m+1}(x) + \frac{2a+2m+1}{2}
(C^{2a+1}_{2m}(x)-mC^{2a+1}_{2m-1}(x))
\ee
The spectral density (5.7) can be written as the sum of two terms
\be
\rho_L(x) &=& x^a e^{-x}\int dy y^a e^{-y} \epsilon(x-y) \left (
\sum_{m=0}^{(n/2)-1}
\frac {1}{r_m} (C^{2a+1}_{2m+1}(x)C^{2a+1}_{2m}(y)-
C^{2a+1}_{2m}(x)C^{2a+1}_{2m+1}(y))\right .  \nonumber\\
&-&
\left . \sum_{m=0}^{(n/2)-1}
\frac {m(2a+2m+1)}{2r_m} (C^{2a+1}_{2m-1}(x)C^{2a+1}_{2m}(y)-
C^{2a+1}_{2m}(x)C^{2a+1}_{2m-1}(y)) \right ).\nonumber\\
\ee
The two sums can be combined into a single sum.
This expression can be simplified further by applying the inverse
Christoffel-Darboux formula (see \cite{GRADSHTEYN-RYZHIK-1980})
and collecting all terms. We find
\be
\rho_L(x) =  x^a e^{-x} \int_0^\infty y^a e^{-y} \frac{|x-y|}2 \sum_{m=0}^{n-2}
\frac{n-(m+1)}{h_m^{2a+1}}C_m^{2a+1}(x)C_m^{2a+1}(y)
\ee
In order to take the limit $n\rightarrow\infty$
the $n$-dependence has to be made more explicit. To achieve this, we apply the
Christoffel-Darboux formula once again to the term proportional to $n$, and,
after using the identity,
\be
C_m^{2a+1} = \frac 1{m+1} \frac d{dx} C_{m+1}^{2a},
\ee
to the term proportional to $(m+1)$. The term proportional to $n$ cancels
against one of the terms obtained from the differentiation with respect to
$x$ and $y$. We finally obtain
\be
\rho_L(x) &=&
\frac {2^{2a} n!}{\Gamma(n+2a)} x^a e^{-x} \int_0^\infty dy y^a e^{-y}
\epsilon(x-y)
\left ( \frac{L_n^{2a}(2x) L_n^{2a-1}(2y)-
L_n^{2a-1}(2x) L_n^{2a}(2y)}{(x-y)^2} \right .\nonumber\\&+& \left .\frac{
L_n^{2a-1}(2x) L_n^{2a+1}(2y)+ L_n^{2a+1}(2x) L_n^{2a-1}(2y)-2L_n^{2a}(2x)
L_n^{2a}(2y)}{x-y}\right ).\nonumber\\
\ee
This sum can be written more compactly as an integral over the unitary kernel
$K(2x,2y)$
\be
\rho_L(x) = x^a e^{-x} \int_0^\infty dy y^a e^{-y}
\epsilon(x-y) (\frac{d}{dy} -\frac{d}{dx} )K(2x,2y),
\ee
with the kernel $K(2x,2y)$ defined by
\be
K(2x,2y) = \frac {2^{2a} n!}{\Gamma(n+2a)}
\frac{L_{n-1}^{2a}(2x) L_n^{2a}(2y)-L_n^{2a}(2x) L_{n-1}^{2a}(2y)}{2x-2y}.
\ee
This kernel was first considered by Fox and Kahn \cite{FOX-KAHN-1964},
and was studied in great detail by Bronk \cite{BRONK-1965}.
The relation (5.21)  shows an intimate
and, an as yet not well understood, relation between the unitary kernel
and the orthogonal spectral density.

\vskip 1.5cm
\renewcommand{\theequation}{6.\arabic{equation}}
\setcounter{equation}{0}
\centerline{\bf 6. The microscopic limit}
\vskip 0.5 cm

In this section we derive the microscopic limit of the spectral density.
First, we express the parameter $\sigma$ in the mean level density.
An expression suitable for the analysis of the spectral density many level
spacings away from the origin but yet far from the edge of the semi-circle
is obtained by commuting the $\epsilon$-function through the derivative
operators.  The result can then be written as the sum of the chGUE spectral
density plus a remaining oscillatory term:
\be
\rho_L(x) = 2 x^{2a} e^{-2x} K(2x,2x)+
x^a e^{-x} \int_0^\infty dy y^a e^{-y}
(\frac{d}{dy} -\frac{d}{dx} )\epsilon(x-y) K(2x,2y).
\ee
Because
\be
2\int_0^\infty  dx  x^{2a} e^{-2x}
K(2x,2x) = n,
\ee
the second term does not contribute to the total number of levels. As is well
known for the chGUE the large $n$-limit of the first term is a semicircle.
In the normalization (4.2) of the distribution of the matrix elements, the
average level density does not depend on $\beta$ (see appendix A).
The second term in (6.1) therefore does not contribute to the average level
density. From
the asymptotic formula for the Laguerre polynomials (see eq. (C.2)),
it follows that in the thermodynamic limit ($n \gg 1$)
\be
\rho_L(x) \sim \frac 2{\pi} \frac n{\sqrt{2nx}},
\ee
for $x\rightarrow 0$. The large-$n$ limit of  the
spectral density (4.6) at $\lambda = 0$
is then given by
\be
\rho(\lambda=0) = \rho_L(x) \frac {dx}{d\lambda} = \frac {N}{\pi\sigma}.
\ee
According to Banks-Casher formula \cite{BANKS-CASHER-1980},
we can identify the parameter $\sigma$ as
\be
\sigma = \frac 1\Sigma,
\ee
where $\Sigma$ is the chiral condensate.

The microscopic limit of the spectral density is given by
\be
\rho_S(z) = \lim_{N\rightarrow \infty} \rho(\lambda = \frac zN) \,\frac
{d\lambda}{dz} = \lim_{N\rightarrow\infty} \rho_L(x = \frac{z^2}{8n\sigma^2})
\,\frac {dx}{dz}.
\ee
In order to evaluate this limit we start from expression
(5.20) for the spectral density and use the microscopic variables $z$ and $w$
defined by
\be
x= \frac {z^2}{8n\sigma^2},\quad{\rm and} \quad y = \frac{w^2}{8n\sigma^2},
\ee
where $z$ is related to the original eigenvalues by $z = \lambda N$.

The term in the integral
proportional to $L_n^{2a+1} L_n^{2a-1}$ does not satisfy the
conditions necessary to
interchange the limit and the integration.
However, if we add and subtract the term
\be
\frac {2^{2a-1} n!}{\Gamma(n+2a)} x^a e^{-x}
\int_0^\infty dy y^{a-1} e^{-y}
 L_n^{2a+1}(2x) L_n^{2a-1}(2y),
\ee
to the integral, it can be proved by dominated convergence that
in the subtracted integral the limit $n\rightarrow\infty$ can
be taken before integration. The integral (6.8) has to be performed
exactly first. The result is (see appendix C)
\be
\lim_{n\rightarrow\infty} n^{-a+1}
\int_0^\infty dy y^{a-1} e^{-y} L_n^{2a+1}(2y) = 2^{-a+1}.
\ee
{}From the asymptotic formula for the Laguerre polynomials (C.3) it then
follows that
the microscopic limit of the spectral density is given by
\be
\rho_S(z) = \frac {\Sigma}{4} J_{2a+1}(z{\Sigma}) &+& \frac {\Sigma}{2}
\int_0^\infty dw (zw)^{2a+1} \epsilon(z-w)
\left ( \frac 1w \frac d{dw} - \frac 1z \frac d{dz}\right )\nonumber \\
&\times&
\frac{wJ_{2a}(z{\Sigma})J_{2a-1}( w{\Sigma})
-zJ_{2a-1}(z{\Sigma})J_{2a}(w{\Sigma})}{(zw)^{2a}(z^2-w^2)}.
\ee
Note that half of the subtracted term is again reabsorbed into the integral.
To achieve this we have used the identity
\be
\frac {\Sigma}{4} J_{2a+1}(z{\Sigma}) = -\frac{\Sigma}{4}\int_0^\infty dw
J_{2a-1}(w\Sigma) z^{2a} \frac {d}{dz} (z^{-2a} J_{2a}(\Sigma z)).
\ee
In eq (6.10) we have expressed the microscopic spectral density in terms
of an integral over the Bessel kernel.  This kernel was studied extensively
by Widom and Tracy \cite{WIDOM-TRACY-1992}.

The leading term in the small $z$-expansion is obtained by
replacing $z^2-w^2\rightarrow -w^2$ and $\epsilon(z-w)\rightarrow -\frac 12$.
Then all integrals can be performed exactly, and after a cancellation
only the integral (6.11) and the first term in (6.10) contribute
to leading order. This results in
\be
\rho(z) \sim \frac {\Sigma}{2\Gamma(2a+2)}
\left (\frac {z\Sigma}{2}\right )^{2a+1}.
\ee

The simplest spectral sum rule is given by
\be
\int_0^\infty dz\frac{\rho_S(z)}{z^2} = \frac {\Sigma^2}{8(N_f+\frac
{\nu-1}2)}.
\ee
This sum rule can also be derived from the partition function without
reference to the spectral density \cite{VERBAARSCHOT-1994S}.

In the case of zero flavors ($N_f = 0$) and zero topological charge ($\nu =0$)
a direct numerical simulation of the random matrix model (4.1) is particularly
simple. To convince the reader that (6.10) is correct we show in Fig. 1
a histogram obtained from the diagonalization of 10,000 $128\times 128$
matrices
(dashed curve) and the exact microscopic limit (6.10) for $a=0$ (full line).
Perfect agreement is no exaggeration in this case. In Fig. 2 we show the
results of the microscopic spectral density for $N_f =1,\, 2$ and 3.
We observe much less oscillations than in the chGUE case. This agrees with
general property known from the classical random matrix ensembles that
spectra of real matrices are much less rigid than spectra of complex matrices.
Therefore, the variation of each level about its average position,
called level motion, is much larger resulting in the (almost)
absence of oscillations.

\vskip 1.5cm
\renewcommand{\theequation}{7.\arabic{equation}}
\setcounter{equation}{0}
\centerline{\bf 7. Conclusions}
\vskip 0.5 cm

In this work we
have studied the spectrum of the
QCD Dirac operator for an $SU(2)$-valued gauge field in the fundamental
representation. This case differs from any other nonabelian gauge group
in the fundamental representation by the reality constraints of
the eigenfunctions: it is possible to choose
a basis in which the matrix elements of the Dirac operator are real.
As we have argued before for the chGUE case,
the correlations between the eigenvalues
of the QCD Dirac operator near zero virtuality are insensitive to the
detailed dynamics of the system and can be
described by a random matrix model with as only its symmetries
as input. There is no reason to believe that the present case is different.
However, the appropriate random matrix ensemble not only has
the chiral structure of QCD but also has real matrix elements.
For this reason, it will be called the chiral orthogonal ensemble,
abbreviated by chGOE.

Because of the spontaneous breaking of chiral symmetry, the spectral density
near zero is $\sim V_4$. This property allows us to define a limit
$V_4\rightarrow\infty$ of the spectral density in which
the eigenvalues are at the same time rescaled by a factor $V_4$.
The resulting spectral density, called the microscopic
spectral density $\rho_S(z)$, is insensitive to the dynamics of the system.
It is a universal function that is entirely determined by the symmetries
of the Dirac operator.

In this paper we have evaluated $\rho_S(z)$ for the case
of $SU(2)$ in which the Dirac operator is real.
This turned out to be much more difficult than for a complex
Dirac operator, a well known property of random matrix theory.
The spectral density in the present case differs from the chGUE case by the
absence of strong oscillations. As is also the case for the classical
random matrix ensembles, the spectrum of a complex matrix is much more rigid
than the spectrum of a real matrix. This can be made more quantitative
in terms of the so called level motion which turns out to be much larger
for the chGOE than for the chGUE.

The microscopic spectral density is generating function for
the Leutwyler-Smilga sum rules. The sum rules for for $N_c = 2$ have not
been obtained before. In view of the fact that the corresponding static
effective field theory involves both baryons (which are bosons in this case)
and mesons \cite{DIAKONOV-PETROV-1993},
it is not surprising that the results are different from
for other nonabelian gauge groups in the fundamental representation.
It would be interesting to derive  the sum rules from
the static limit of the effective field theory also in this case.
In the case of one flavor the effective theory for $N_c = 2$ and other
nonabelian gauge groups coincides. Although the spectral density
is different, all spectral sum rules for
$N_f =1$ turn out to be the same for the chGOE and chGUE. Further work
to clarify this issue is in progress \cite{VERBAARSCHOT-1994S}.

\vskip 1.5cm
\renewcommand{\theequation}{A.\arabic{equation}}
\setcounter{equation}{0}
\noindent
\centerline{\bf Appendix A}
\vskip 0.5cm

The joint eigenvalue density valid for both the chGOE ($\beta= 1$) and the
chGUE
($\beta =2$) is given by
\be
\rho_\beta(\lambda_1, \cdots, \lambda_n) = C_{\beta, n}
\prod_{k,l} |\lambda_k^2 -\lambda_l^2|^\beta \prod_{k}
\lambda_k^{\alpha}
\exp({-\frac{n\beta\Sigma^2}{2} \sum_k \lambda^2_k}).
\ee
where $C_{n,\beta}$ are normalization constants and
$ \alpha =(2N_f +\beta\nu+\beta -1)$.\\ \noindent
The normalization integral
$\int d\lambda_1 \cdots d\lambda_n \rho_\beta(\lambda_1, \cdots, \lambda_n)$
 can be approximated by
\be
\exp\left[ \beta \int d\lambda d\lambda'
\log|\lambda^2-{\lambda'}^2| \bar\rho(\lambda) \bar\rho(\lambda')+
\int d\lambda \bar\rho(\lambda)\left (
-\frac{n\beta}{2\sigma^2} \lambda^2
+ \alpha\log\lambda + \mu\bar\rho(\lambda)\right
)\right],\nonumber\\
\ee
where the average level density $\bar \rho(\lambda)$ satisfies a 'mean-field'
equation obtained by minimizing the exponent:
\be
2\beta\int d\lambda' \log|\lambda^2-{\lambda'}^2| \bar\rho(\lambda')
-\frac{n\beta}{2\sigma^2}\lambda^2 + \alpha\log\lambda + \mu = 0.
\ee
The normalization of the level density is introduced via a Lagrange
multiplier.
By differentiation with respect to $\lambda^2$ we obtain the principal value
equation
\be
2\beta P\int \frac {d\lambda'}{\lambda^2 - {\lambda'}^2} \bar\rho(\lambda') -
-\frac{n\beta}{2\sigma^2}+ \frac {\alpha}{2\lambda^2} = 0.
\ee
Since $\int \bar\rho(\lambda)d\lambda = n$ the third term is subleading for
$n\rightarrow \infty$ (in agreement with general properties of
topological fluctuations we have $\nu \ll n$). Consequently, the
'mean-field' equation for $\bar\rho$ does not depend on $\beta$.
A more elaborate discussion of this argument for the classical random
matrix ensembles
can be found in the book by Mehta \cite{MEHTA-1991}.
\vskip 1.5cm
\renewcommand{\theequation}{B.\arabic{equation}}
\setcounter{equation}{0}
\noindent
\centerline{\bf Appendix B}
\vskip 0.5 cm
In this appendix we calculate the Jacobian of the transformation of the matrix
valued variables $T$ into its eigenvalues and eigenangles using techniques
developed in \cite{HUA-1963}.
For an arbitrary
real $n\times m$ matrix we have
\be
T = {\cal O}_1 \Lambda {\cal O}_2,
\ee
where $\Lambda$ is a positive $n\times m$ diagonal matrix and,
the $n\times n$ matrix ${\cal O}_1$ and the $m\times m$ matrix ${\cal O}_2$
are orthogonal (for definiteness $m>n$).
By differentiation of (B.1) it can be shown that that the
variation $dT$ can be expressed in variations $\delta {\cal O}_i$
of ${\cal O}_i$ near the identity
\be
{\cal O}_1^{-1} dT {\cal O}_2 &=& \delta {\cal O}_1 \Lambda + d\Lambda -\Lambda
\delta{\cal O}_2,\\
{\cal O}_2^{-1} d\tilde T {\cal O}_1 &=& -\Lambda\delta {\cal O}_1 + d\Lambda
+\delta{\cal O}_2\Lambda,
\ee
where we have introduced $\delta {\cal O}_i = {\cal O}_i^{-1}d{\cal O}_i$.
Note that the matrices $\delta {\cal O}_i$ are anti-symmetric, and, in
particular the diagonal matrix elements are zero. For the invariant distance we
find
\be
{\rm Tr} dT d\tilde T = \sum_k (d\Lambda_k)^2 &+& \sum_{k<l}^n \frac 12\left[ (
\delta {\cal O}_1 -
\delta {\cal O}_2)_{kl}\right ]^2 (\lambda_k +\lambda_l)^2
+\sum_{k<l}^n \frac 12\left[ (
\delta {\cal O}_1 +\delta {\cal O}_2)_{kl}\right ]^2 (\lambda_k -\lambda_l)^2
\nonumber\\
&+& \sum_{k=1}^n \sum_{l=n+1}^m [(\delta {\cal O}_1)_{kl}]^2 \lambda_k^2.
\ee
This allows us to read off the Jacobian of the transformation to the variables
$ d\Lambda_k$ and $(\delta {\cal O}_1 \pm\delta {\cal O}_2)/\sqrt 2$ from
the Lam\'e-coefficients:
\be
J = \prod_{k<l}^n (\lambda_k -\lambda_l)^2 \prod_k^n \lambda_n^{|m-n|}.
\ee
Note that the total powers of lambda can be verified by a dimensional argument.

\vskip 1.5cm
\renewcommand{\theequation}{C.\arabic{equation}}
\setcounter{equation}{0}
\noindent
\centerline{\bf Appendix C}
\vskip 0.5 cm
In this appendix we consider the large $n$ limit of integrals of the type
\be
I_n^{a,k}= n^{k+1-a}\int_0^\infty L_n^{2a-k}(2y) y^a e^{-y} dy,
\ee
where $k$ is an integer, and $n$ is even. From the asymptotic expansion
for $n\rightarrow\infty$ of the generalized Laguerre polynomials (see
\cite{GRADSHTEYN-RYZHIK-1980} for this and other properties of
the generalized Laguerre polynomials used below), among others,
\be
L_n^\alpha(x) \sim \frac 1{\sqrt \pi} e^{\frac x2} x^{-\frac\alpha{2} -\frac
14} n ^{\frac \alpha{2} -\frac 14} \cos(2\sqrt{nx} -\frac {\alpha\pi}{4}
-\frac{\pi}{4}),
\ee
it is clear that for $k < -3/2$ and
$a > -1$ the absolute value of the integrand can be majorated by an integrable
function. In that case we can, after introducing a new integration variable by
$ y = w^2/2n$, interchange the limit $n \rightarrow \infty$
and the integration over $w$. Using the asymptotic result
\be
\lim_{n\rightarrow \infty} L_n^\alpha(\frac{w^2}{n}) \sim n^{\alpha}
w^{-\alpha} J_\alpha(2w),
\ee
with $J_\alpha$ a Bessel function, we find that
\be
\lim_{n\rightarrow \infty} I_n^{a,k} = 2^{-a}\int dw w^{k+1} J_{2a-k}(2w).
\ee
This integral can be evaluated analytically, resulting in
\be
\lim_{n\rightarrow \infty} I_n^{a,k} = 2^{-a-1}
\frac{\Gamma(a+1)}{\Gamma(a-k)}.
\ee

For $-\frac 32 < k < -\frac 12$ (since $k$ is an integer: $k = -1$)
and $a > -1$ the integral still converges
but no longer satisfies the conditions necessary to interchange
the limit and the integral.
In this case there is an important
contribution from the region around the largest zero of $L_n^\alpha$.
However, for $k = -1$, it is particularly simple to evaluate the
integral for any  finite value of $n$. From the recursion relation
\be
L_n^{2\alpha+1}(2y) = \sum_{m=0}^n L_m^{\alpha}(y)L_{n-m}^{\alpha}(y)
\ee
we can reduce the integral to a normalization integral for the Laguerre
polynomials resulting in
\be
I_n^{a,-1} = n^{-a} \frac {\Gamma(a+ \frac n2 + 1)}{\Gamma(\frac n2+1)}.
\ee
The asymptotic limit follows immediately from Stirlings formula
\be
\lim_{n\rightarrow \infty} I_n^{a,-1} = 2^{-a},
\ee
which is a factor 2 bigger than the result given in (C.5). We conclude
that half of the contribution to this integral is from the region near the
largest zero of the Laguerre polynomial.
\vfill
\eject
\newpage

\vglue 0.6cm
{\bf \noindent  Acknowledgements \hfil}
\vglue 0.4cm
 The reported work was partially supported by the US DOE grant
DE-FG-88ER40388. The author wants to thank A. Smilga for pointing out
the special status of $N_f = 1$. P.~Kahn is acknowledged for useful
discussions.

\vfill
\eject
\newpage
\setlength{\baselineskip}{15pt}

\bibliographystyle{aip}
\vfill
\eject
\newpage
\noindent
{\bf Figure Captions}
\vskip 0.5cm
\noindent
Fig. 1. The microscopic spectral density $\rho_S(z)$ versus $z$ for
$N_f = 0$, $\nu = 0$ and $\Sigma = 2$.
The full line represents the exact analytical result
and the cashed curve shows data obtained by diagonalizing 10,000 random
matrices distributed according to (4.1).
\vskip 0.5 cm
\noindent
Fig. 2. The microscopic spectral density $\rho_S(z)$ versus $z$ for
$N_f = 1$ (dotted curve), $N_f = 2$ (dashed curve) and $N_f = 3$ (full curve)
all for $\nu = 0$ and $\Sigma = 2$.
\end{document}